\documentclass[aps,prb,twocolumn,superscriptaddress]{revtex4}
\usepackage{natbib}
\usepackage{amsmath}
\usepackage{amssymb}
\usepackage{amsthm}
\usepackage{graphicx}
\usepackage{graphics}
\usepackage{subfigure}
\usepackage{bm}
\usepackage{hyperref}

\newcommand{\abs}[1]{\left\vert#1\right\vert}

\def\beas{\begin{eqnarray*}}
\def\eeas{\end{eqnarray*}}
\def\bea{\begin{eqnarray}}
\def\eea{\end{eqnarray}}
\def\be{\begin{equation}}
\def\ee{\end{equation}}

\begin{document}

\title{The longitudinal resistance of a quantum Hall system
with a density gradient}%

\author{R.\ Ilan}
\affiliation{Department of Condensed Matter Physics, Weizmann
Institute of Science, Rehovot 76100, Israel}
\author{N.\ R.\ Cooper}
\affiliation{T.C.M. Group, Department of Physics, Cavendish
Laboratory, J.J. Thomson Avenue, Cambridge, CB3 0HE, U.K.}
\author{Ady Stern}
\affiliation{Department of Condensed Matter Physics, Weizmann
Institute of Science, Rehovot 76100, Israel}

\date{\today}

\begin{abstract}
Following recent experiments, we consider current flow in two
dimensional electronic systems in the quantum Hall regime where a
gradient in the electron density induces a spatial variation in the
Hall resistivity. Describing the system in terms of a spatially
varying local resistivity tensor, we show that in such a system the
current density is generically non-uniform, with the current being
pushed towards one side of the sample. We show that, for sufficiently
large density gradient, the voltage along that side is determined by
the change of the Hall resistivity, and is independent of the
microscopic longitudinal resistivity, while the voltage on the
opposite side is small and determined by both longitudinal and Hall
resistivities. We solve some particular models in detail, and propose
ways by which the microscopic longitudinal resistivity may be
extracted from measurements of the longitudinal voltage on both sides
of the sample.

\end{abstract}


\maketitle


\section{Introduction}

Recent striking experimental observations of transport in high
mobility two-dimensional electron gases in the quantum Hall regime
have been successfully explained as resulting from density gradients
of as small as $1\%/\mbox{cm}$.\cite{weipan} These gradients
presumably arise from inhomogeneities in the MBE growth.\cite{weipan}
Specifically, the longitudinal voltage drop differs on the two
sides of the sample: on one side the voltage drop is close to
zero; on the other the voltage drop yields a longitudinal
resistance equal to the difference in the {\it Hall} resistivities
corresponding to the densities at the two probes.
Thus, the longitudinal resistance of such samples does not measure
an intrinsic longitudinal resistivity, $\rho_{xx}$, but rather
probes the (spatially varying) Hall resistivity $\rho_{xy}$.

The explanation of these observations given  in
Ref.~\onlinecite{weipan} was based on the edge-state formalism for
transport in quantum Hall systems.
This formalism is capable of describing inhomogeneous
systems consisting of regions of quantized Hall conductance that
are separated by sharp boundaries. However, there are many
questions that it cannot address, such as
how ``sharp'' must the boundary between the quantized Hall regions
be for the edge-state formalism to apply? And
how would the transport properties of a {\it compressible} system,
such as the $\nu=1/2$ composite Fermi liquid,\cite{hlr} be affected by
a density gradient?  In this case the sample is at no point in a
quantum Hall state so cannot be described within the edge-state
formalism.

In this work we employ a more general theoretical approach: we model
the experimental system as an inhomogeneous classical
conductor,\cite{Ruzinstep,st} with a spatially varying local
resistivity tensor set by the inhomogeneous electron density, {\it
with a net density gradient}. Using this model we are able to answer
the questions posed above and analyze the relation between the
observed transport phenomena and the properties of the quantized Hall
effect. As we show below, {\it under quite general circumstances, the
presence of a density gradient results in a non-uniform current
density that leads to a difference in the longitudinal voltage drop on
the two sides of a Hall bar sample.}
We describe the conditions under which the boundary between the
quantum Hall states is ``sharp'' and the edge-state formalism used
in Ref.~\onlinecite{weipan} applies; we also determine corrections
to this behavior.  We discuss in detail the effects of a density
gradient on a compressible state.  In this case we show that, even
in the presence of an {\it unknown} density gradient, the
intrinsic local resistivity $\rho_{xx}$ can in principle be
extracted from resistance measurements in a Hall bar geometry.

There have been many previous studies of quantum Hall systems
using this
approach.\cite{Ruzinstep,st,ResistivityLaw,dykhneruzin,rch,chhr}
These have shown that, for macroscopically homogeneous samples
(when the correlation length of the density inhomogeneities is
small compared to the sample size), the measured longitudinal
resistance can, depending on the experimental circumstances, be
determined by the fluctuations in $\rho_{xy}$ independent of the
intrinsic $\rho_{xx}$:\cite{st,ResistivityLaw,dykhneruzin}
notably, the height of the peak in the longitudinal resistivity
between two quantised Hall states can be equal to the difference in
the Hall resistivities of those
states.\cite{dykhneruzin} However, in all macroscopically
homogeneous systems, the {\it same} longitudinal voltage drop
would appear on the two sides of a Hall bar sample.
While  a density gradient was invoked in Ref.~\onlinecite{chhr} to
account for transport measurements of certain Corbino disk
samples,\cite{rsg} the effects of a density gradient have not been
adequately explored. Other works considered related effects due to
inhomogeneities in the magnetic field and variations in the sample
thickness.\cite{Bruls,Chaplik}
The paper is organized as follows.
After formulating the mathematical problem in
\S\ref{sec:formulate}, in \S\ref{sec:general} we provide general
arguments for the transport properties of a system in which
voltage probes lie in two different quantized Hall states.  We
show how the results of Ref.~\onlinecite{weipan} emerge within our
theoretical approach.  Our conclusions here rely on an assumption
that current remains close to one side of the sample.
Justification for this assumption is provided by
\S\ref{sec:uniform},\S\ref{sec:rectangular} and
\S\ref{sec:dissipative} which contain explicit solutions for
particular geometries:
in \S\ref{sec:uniform} and \S\ref{sec:rectangular} we study a
compressible system with a uniform density gradient; in
\S\ref{sec:dissipative} we calculate the effects of a uniform
dissipative region between the two quantized Hall regions.
Finally, in \S\ref{sec:discuss} we provide discussion and
conclusions.

\section{Formulation of the problem}
\label{sec:formulate}

We consider transport in a two-dimensional sample in the $x-y$
plane. The local current density $\mathbf{j}(\mathbf{r})$ is
related to the local electric field
$\mathbf{E}(\mathbf{r})$\footnote{Strictly speaking $\mathbf{E}$
is (minus) the gradient of the electrochemical potential, and
$\mathbf{j}$ is the transport current density.} via
\begin{eqnarray}
\mathbf{E} & = & \rho_{xx} \mathbf{j} - \rho_{xy}
\mathbf{\hat{z}}\times \mathbf{j}
 \label{rho}
\end{eqnarray}
where the components of the resistivity tensor, $\rho_{ij}$, are also
assumed to be functions of position.   These
variations arise from inhomogeneities in the
underlying carrier density in the quantum Hall samples.

In a steady state, the current density and the electric field must
satisfy
\begin{eqnarray}
&&\mathbf{\nabla}\cdot\mathbf{j}=0 \label{continuity}\\
&&\mathbf{\nabla}\times\mathbf{E}=0 \label{curlE}
\end{eqnarray}
as well as the appropriate boundary conditions.

We can ensure (\ref{curlE}) by defining a scalar potential $\phi$,
such that \be \label{static} \mathbf{E}=-\mathbf{\nabla}\phi. \ee
Similarly, we can ensure (\ref{continuity}), by defining a stream
function $\psi$
 via
\begin{equation}
\mathbf{j}=\mathbf{\hat{z}}\times\mathbf{\nabla}\psi\label{stream}.
\end{equation}
Then, substituting (\ref{rho}) into (\ref{curlE}), the
equation for $\psi$ is
\begin{equation}
 \rho_{xx} \nabla^2\psi
+\mathbf{\nabla}\rho_{xx} \cdot\mathbf{\nabla}\psi
 - \mathbf{\hat{z}} \cdot \mathbf{\nabla}\psi \times \mathbf{\nabla}\rho_{xy}  =
 0.
\label{fullcurl}
\end{equation}

We shall solve equation
(\ref{fullcurl}) using various boundary conditions, and functional
dependences of the local resistivity tensor on position.

\section{General Arguments}

\label{sec:general}

\begin{figure}[!t]
\begin{center}
    \includegraphics[width=70mm]{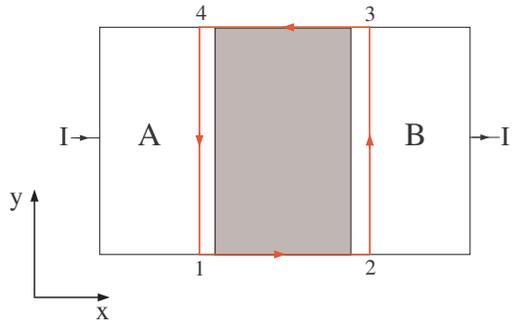}
    \caption{Geometry of the sample considered. A total current $I$
flows from left to right along the length of the sample.  The four
voltage probes lie on the sides of the sample within the two
quantised Hall regions ($A$ and $B$) at the two ends.  The transport
properties of the intervening shaded region are unimportant for the
discussion of \S~\ref{sec:general}.  The contour of integration used
for calculating the dissipation integral in the Appendix is shown. }
\label{contour}
\end{center}
\end{figure}

We begin our discussion by considering the situation in which the
change in density along the sample is sufficient that the two ends
of the sample are in different quantized Hall states, $A$ and $B$.
We assume that these regions exhibit good quantized Hall effects,
{\it i.e.}  that dissipation is negligible $|\rho_{xx}/\rho_{xy}|
\ll 1$. These regions are therefore characterized just by their
Hall resistivities $\rho_{xy}^A$ and $\rho_{xy}^B$.  As
illustrated in Fig.~\ref{contour}, we imagine that the four
voltage probes attached to the sample lie in these regions of
quantized Hall resistivity. For now, the details of the
resistivity in the intervening region (shown shaded in
Fig.~\ref{contour}) are unimportant. Moreover, the precise
geometry is not important: the density gradient could be in any
direction, or indeed could be nonuniform -- we require only that
contacts 1 and 4 are in region $A$, and 2 and 3 in $B$.

A key result of our work is to show that when a current $I$ is
passed
along a sample of this type:\\
(i) the longitudinal voltage drop is larger on one side of the
sample than the other, the side being selected by the sign of
$\Delta\rho_{xy} \equiv \rho_{xy}^A - \rho_{xy}^B$;\\
(ii) under physically realistic conditions for the intervening
region, the smaller of these two voltages can be exponentially
small (in a
manner to be discussed in detail below);\\
(iii) in cases where (ii) applies, the nonzero voltage yields, to
exponential accuracy, a longitudinal resistance
\be R = |\Delta \rho_{xy}| \ee
That is, the apparent longitudinal resistance is determined by the
difference in the {\it Hall} resistivities at the locations of the
two contacts.

These effects are identical to those observed experimentally in
Ref.~\onlinecite{weipan}, and derived there within the edge state
transport formalism.  However, we emphasize that our results
hold in a much more general setting. For example, we shall show
that the two quantized Hall fluids may be joined by a finite
region of a compressible quantum Hall fluid where the resistivity
tensor is not quantized. Furthermore, we shall be able to account
for departures from the limiting behavior of results (ii) and
(iii).

First, we prove result (i) by adapting an approach of
Ref.~\onlinecite{rch} to the present geometry. As described in the
Appendix, by making use of the fact that the regions $A$ and $B$
have negligible dissipation one finds that, {\it independent of
the spatial distribution of the resistivity between these two
regions}, the net power dissipated within the sample is \be
\label{eq:q} Q = \frac{1}{2}\left(\rho_{xy}^{A} -
\rho_{xy}^{B}\right)^{-1} \left( V_{43}^2 - V_{12}^2 \right) \ee
where
\begin{eqnarray}\label{volatage}
V_{12} & \equiv & \int_{x_{1}}^{x_{2}}\mathbf{E}\cdot d\mathbf{l}=\phi(x_{1})-\phi(x_{2})\\
V_{43} & \equiv & \int_{x_{4}}^{x_{3}}\mathbf{E}\cdot
d\mathbf{l}=\phi(x_{4})-\phi(x_{3})
\end{eqnarray}
are the longitudinal voltage drops on the two sides of the sample.
Since the dissipated power must be positive, $Q>0$, Eq.(\ref{eq:q})
shows that $|V_{43}|> |V_{12}|$ if $\rho_{xy}^{A}
> \rho_{xy}^{B}$ (and vice versa).  This proves result (i); it
also identifies which of the voltage drops is the larger for a
given sign of $\Delta \rho_{xy}$.

The demonstration of result (ii) will be discussed in the
following sections. For now, we note that result (iii) is a direct
consequence of (ii) in the geometry described. Let us assume that
the necessary conditions have been met that (ii) holds.  For
simplicity, we imagine that the sample has been oriented such that
$\rho_{xy}^{A} > \rho_{xy}^{B}$ in which case $|V_{43}| >
|V_{12}|$. Then it is $V_{12}$ that is exponentially small and which
we now set to zero. We can now evaluate the potential drop
$V_{43}$ by integrating the electric field from 4 to 1 to 2 to 3.
Using the facts that the paths $4\to 1$ and $2\to 3$ are in
quantized Hall states, and that $V_{12}=0$, one finds \bea V_{43}&
=& I \left(\rho_{xy}^{A} - \rho_{xy}^{B} \right)\\ R_{43} \equiv
\frac{V_{43}}{I}& = & \rho_{xy}^{A} - \rho_{xy}^{B} =
\Delta\rho_{xy} \eea This is result (iii), which we now see as a
consequence of (ii).

In this section we have asserted that the smaller voltage drop can
be exponentially small, as in result (ii). In the following
sections we describe the current distributions in particular
(physically relevant) situations where exact results can be
obtained, and which do show that the smaller voltage drop is,
indeed, exponentially small.

\section{Transport in a Hall Bar with a Uniform Gradient in the Hall Resistivity}

\label{sec:uniform}

We first consider the situation of a long Hall bar aligned with
the $x$-axis, and with edges at $y=0,w$.  We shall discuss the
current distribution in a region that is far from the current
contacts.  We consider the situation in which $\rho_{xx}$ is
independent of position, and $\rho_{xy}$ has a constant gradient,
$\mathbf{\nabla}\rho_{xy}$ (which can be in an arbitrary
direction).  As is discussed in more detail in
\S\ref{sec:discuss}, this model can be valid in a sample at a
density close to a compressible state and with a Hall angle close
to $90^\circ$ ($\rho_{xx}\ll |\rho_{xy}|$). For example, these
conditions can be met in a high-quality sample at a density close
to the $\nu=1/2$ Composite Fermion liquid state.\cite{hlr}

\subsection{Current distribution}
\label{Currentdistribution}

Under the  conditions stated above, Eq. (\ref{fullcurl}) reduces
to
\begin{equation}
\rho_{xx}\nabla^2\psi-\frac{\partial\psi}{\partial
x}\frac{\partial \rho_{xy}}{\partial y} +
\frac{\partial\psi}{\partial y}\frac{\partial \rho_{xy}}{\partial
x}=0 \,. \label{eq:simple}
\end{equation}
The constant parameters $\rho_{xx}$ and $\mathbf{\nabla}\rho_{xy}$
introduce two new lengthscales
\begin{eqnarray}
\label{eq:lengthx}
 \ell_x  & \equiv  & \rho_{xx} \left(\frac{\partial \rho_{xy}}{\partial
x}\right)^{-1}\\
 \ell_y  & \equiv  & \rho_{xx} \left(\frac{\partial \rho_{xy}}{\partial
y}\right)^{-1} \label{eq:lengthy}
\end{eqnarray}
which characterise the spatial variation of the resistivity
tensor: $\abs{\ell_{x}}$ ($\abs{\ell_y}$) is the distance over
which $\rho_{xy}$ changes by $\rho_{xx}$ in the $x$ ($y$)
direction.

There is a simple solution to Eq.(\ref{eq:simple}) that is
consistent with the boundary conditions that $j_{y}=0$ at $y=0, w$
on the top and bottom sides of the Hall bar. It is sufficient to
set $\psi(\mathbf{r}) = \psi(y)$, such that $j_{y} =
\frac{\partial \psi}{\partial x} = 0$ everywhere. Then
Eq.(\ref{eq:simple}) becomes
$$ \frac{\partial^2 \psi }{\partial y^2} = -
\frac{1}{\rho_{xx}}\frac{\partial \rho_{xy}}{\partial x}
\frac{\partial \psi}{\partial y},$$ {\it i.e.}
$$ \frac{\partial j_x }{\partial y} = -
\frac{1}{\rho_{xx}}\frac{\partial \rho_{xy}}{\partial x} j_x.$$ The
solution is
\begin{equation}
\label{eq:sol} j_x = \frac{I}{\ell_x} \frac{
e^{-y/\ell_x}}{(1-e^{-w/\ell_x})} \quad , \quad j_y=0
\end{equation}
where $I \equiv \int_0^{w} j_x \,dy$ is the total current flowing
along the sample and ${\ell_x}$ is the characteristic
lengthscale (\ref{eq:lengthx}). The solution describes a current
that is concentrated within a distance of order $|\ell_x|$ of
one edge of the sample. Which edge this is depends
on the sign of $\ell_x$ ({\it i.e.} the sign of $\frac{\partial
\rho_{xy}}{\partial x}$ since $\rho_{xx}$ is  positive). For
$\ell_x > 0$, the current is concentrated
close to $y=0$; for
$\ell_x < 0$, the current is concentrated
close to $y=w$.

From this solution, it is straightforward to determine the resistances
that would be obtained in a four-terminal measurement on the Hall
bar. We assume that the voltage probes are a distance $L$ apart, so
that, depending on the side of the sample on which these probes are
placed, the two longitudinal resistances are
\begin{eqnarray}
\label{eq:rl0}
R_{12} & = & \frac{E_x(y=0)L}{I}  =  \frac{L}{\ell_x}\rho_{xx}\frac{1}{1-e^{-w/\ell_x}}\\
R_{43}  & = & \frac{E_x(y=w)L}{I}
     =  R_{12}(0)e^{-w/\ell_x}
\label{eq:rlw}
\end{eqnarray}
(We label the voltage probes in the same way as in
Fig.~\ref{contour}. Note, however, that here we do not consider an
inhomogeneous system in which there are two regions of quantized
Hall states: the gradient in Hall resistivity is assumed to be
uniform and constant over the whole bar.)

In the limit $w\ll |\ell_x|$, the current is spread uniformly over
the width of the sample and we recover the results for a
homogeneous system: the longitudinal resistance is the same on
each side of the sample and set by the bulk resistivity
$\rho_{xx}$ and a geometrical factor
$$R_{12} = R_{43} = \rho_{xx}\frac{L}{w}\,.$$

In the limit $w\gg |\ell_x|$, the current is strongly confined to one
side of the sample. Assuming that $\ell_x
 < 0$ so the current is close to the $y=w$ side, we find
\begin{eqnarray*}
R_{12} & = & 0\\
R_{43} & = & L \left|\frac{\partial \rho_{xy}}{\partial x}\right| = |\Delta \rho_{xy}|
\end{eqnarray*}
(For $\ell_x > 0$ the current is pushed
to the other side, so $R_{43} =0$ and $R_{12}=|\Delta \rho_{xy}|$.)
For $w\gg |\ell_x|$, on the side of the sample where the current is
concentrated, the longitudinal resist{ance drop between the two
voltage probes is set by the difference $|\Delta \rho_{xy}|$ of the
Hall resistivities at these two points; on the other side the
longitudinal potential vanishes.  This result is identical to that for
the case in which the contacts lie in two quantum Hall fluids, as
discussed in \S\ref{sec:general}.  We emphasize, however, that the
result described here applies to the very different situation of a
{\it compressible} system in which $\rho_{xy}$ varies smoothly in
space.

Nevertheless, this solution provides an explicit example of a
situation in which the voltage drop on one side of the sample is
exponentially small.  If the region between the quantized Hall
phases of Fig.~\ref{contour} were to consist of a region of linearly
varying Hall resistivity, as described by the model studied in this
section, the current would be tightly confined to one side of the
sample within a lengthscale $\ell_x$.  The voltage drop on the other
side of the sample would be exponentially small, and the arguments
of \S~\ref{sec:general} can be applied.  In this case, we see that
the parameter controlling the exponential suppression is the ratio,
$w/|\ell_x|$, of the width of the Hall bar $w$ to the lengthscale
$|\ell_x|$.

\subsection{Extracting $\rho_{xx}$}

It is interesting to note that even in the presence of an {\it
unknown} gradient $\mathbf{\nabla} \rho_{xy}$, the bulk
resistivity $\rho_{xx}$ can, in principle, be extracted from the
two measured resistances $R_{12}, R_{43}$ in this Hall bar
geometry, using the results (\ref{eq:rl0},\ref{eq:rlw}). (It is
important that the conditions for validity of these results apply.
These conditions are discussed further in \S\ref{sec:discuss}.)
Our solution shows that the component of the gradient of the Hall
resistivity perpendicular to the Hall bar, $\partial
\rho_{xy}/\partial y$, is irrelevant for the long Hall bar (but
see \S\ref{sec:rectangular} for finite samples). The component of
the gradient for the Hall resistivity along the Hall bar can be
deduced from (\ref{eq:lengthx},\ref{eq:rl0},\ref{eq:rlw}) to be
\begin{equation}
\frac{\partial \rho_{xy}}{\partial x} = \frac{R_{12}-R_{43}}{L}
\end{equation}
Moreover, simple manipulations show that
\begin{equation}
\rho_{xx} = \frac{w}{L} \frac{R_{12}-R_{43}}{\ln
\left[R_{12}/R_{43}\right]}. \label{eq:rhoxx}
\end{equation}
For systems in which $|\ell_x|$ is small compared to the sample
width $w$, the accuracy with which $\rho_{xx}$ can be extracted
becomes limited by the accuracy with which the smaller resistance
can be measured. In this case, in order to extract
$\rho_{xx}$ reliably would require measurements on narrower
samples, in which $w/|\ell_x|$ is of order one so that
$R_{12}/R_{43}$ is not too small or too large.

The results of this section rely on the geometry of a long Hall bar,
such that in the vicinity of the voltage probes the current
distribution is independent of position along the bar.  We now turn to
discuss the resistance that would be measured in rectangular
geometries, with the current contacts at the corners.


\begin{figure}[!t]
\begin{center}
    \includegraphics[width=45mm]{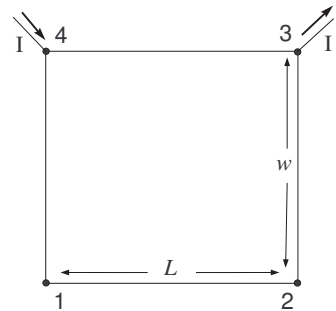}\\
    \caption{Illustration of the system described in \S~\ref{sec:rectangular}. Contacts are numbered
    $1-4$, such that contact $1$ is at the origin. }
\label{description}
\end{center}
\end{figure}

\section{Transport in Rectangular samples with a Uniform Gradient in the Hall Resistivity}
\label{sec:rectangular}

We now consider again a system in which there is a constant
gradient to $\rho_{xy}$, and $\rho_{xx}$ is independent of
position. However, now we study a sample which is a rectangle of
length $L$ and width $w$, and in which current is injected and
extracted at the corners, as described in Fig.~\ref{description}.
This is representative of the geometry of the samples used in
Ref.~\onlinecite{weipan}. The important new aspect, as compared to
the case of the long Hall bar, is that the current is injected
into the sample in a nonuniform profile and we describe the
current distribution in the whole sample, including the bending of
the current flow in the vicinity of the current contacts.

We therefore aim
to solve Eq. (\ref{eq:simple}) again, with boundary
conditions that correspond to two current contacts at the two corners $(0,w)$ and $(L,w)$.
These boundary conditions are
\begin{eqnarray}\nonumber
&&\psi(x,y=0)=\psi(x=0,y)=\psi(x=L,y)=C_{1}\\
&&\psi(x,y=w)=C_{2},
\end{eqnarray}
where $C_{1}, C_{2}$ are constants. Without loss of generality, we
can choose $C_{1}=0$, $C_{2}=-I$. The solution to (\ref{eq:simple})
with these boundary conditions is
\begin{widetext}
\begin{eqnarray}
\psi =
\sum_{m=0}^{\infty}-I\frac{2m\pi(1-(-1)^{m}e^{-\frac{L}{2\ell_y}})}{(\frac{L}{2\ell_y})^{2}+m^2\pi^2}\exp{(\frac{x}{2\ell_y}-\frac{(y-w)}{2\ell_x})}
\frac{\sinh(\frac{1}{2}\sqrt{(\frac{1}{\ell_x})^2+4\lambda_{m}}y)}{\sinh(\frac{1}{2}\sqrt{(\frac{1}{\ell_x})^2+4\lambda_{m}}w)}\sin(\frac{m\pi
x}{L}),\label{full solution}
\end{eqnarray}
\end{widetext}
where
\begin{eqnarray}
\lambda_{m}=\frac{m^2\pi^2}{L^2}+\left(\frac{1}{2\ell_y}\right)^2.
\end{eqnarray}
We note that there are four length scales relevant to this
problem: $w, L,\abs{\ell_{y}}, \abs{\ell_{x}}$.

In the following we examine first the case where the gradient of
$\rho_{xy}$ is in the $x$--direction ($\ell_y=0$). We calculate the
current profile and the longitudinal voltages on the two sides $y=0,w$
for the cases of positive and negative $\ell_x$. When $\abs{\ell_x}\ll
L,w$, we find that the current flows primarily along one of the sides,
that is determined by the sign of $\ell_x$. The voltages on the two
sides are again different, but the presence of the contacts require a
certain care in analyzing the voltage drops. Then, we examine the case
of $\ell_x=0$, where the gradient of $\rho_{xy}$ is in the
$y$--direction. In that case, the effect of the density gradient on
the current flow is more minor.

\subsection{Gradient in the $x$-direction}

When $\partial\rho_{xy}/\partial y=0$ Eq. (\ref{full solution})
becomes
\begin{eqnarray}
\psi = \sum_{m(odd)}-I\frac{4}{m\pi}
\frac{\sinh(\frac{1}{2}\sqrt{4\lambda_{m}+\frac{1}{\ell_{x}^2}}y)}{\sinh(\frac{1}{2}\sqrt{4\lambda_{m}+\frac{1}{\ell_{x}^2}}w)}\times\\\nonumber\times\sin(\frac{m\pi
x}{L})\exp{-\frac{(y-w)}{2\ell_{x}}}.\label{solution1}
\end{eqnarray}
The current components can be extracted using Eq.(\ref{stream}).

In the limit where $\ell_x\ll L,w$ the voltage drop along the
$y=0$ side is
\begin{eqnarray}\nonumber
&V_{12}&= \rho_{xx}\sum_{m(odd)}I\frac{8L}{m^2\pi^2\abs{\ell_x}}
\sqrt{4m^2\pi^2\left(\frac{\ell_x}{L}\right)^2+1}\times\\&&\times\exp{(\frac{w}{2\ell_{x}}-\frac{1}{2}\sqrt{4m^2\pi^2\left(\frac{\ell_x}{L}\right)^2+1}\frac{w}{\abs{\ell_x}})}.
\end{eqnarray}
When $\ell_x<0$, this voltage drop is exponentially small, as
$e^{-w/\ell_x}$. When $\ell_x>0$, the first few terms in the sum
(those with $m\ll L/|\ell_{x}|$) yield a significant finite
contribution which is proportional to $\rho_{xx}\frac{L}{\ell_x}$,
and the first correction is smaller by a factor of $\ell_x/L$.
Therefore we have
\begin{equation}
R_{12}\cong\left\{
\begin{array}{ll}
0 + {\cal O}(e^{-w/\ell_x}) & \quad\ell_x<0 \\
\rho_{xx}\frac{L}{\ell_x}=\Delta\rho_{xy} + {\cal O}(\rho_{xx}) & \quad\ell_x>0 \\
\end{array}
\right..
\end{equation}

The voltage drop on the $y=w$ edge is
\begin{eqnarray}\nonumber\label{diverge sum}
V_{43}=-\rho_{xx}\sum_{m(odd)}&&I\frac{8L}{m^2\pi^2}\left(\frac{1}{2\ell_x}-\frac{1}{2}
\sqrt{4\frac{m^2\pi^2}{L^2}+\frac{1}{\ell_{x}^2}}\times\right.\\
&&\left.\times\coth(\frac{1}{2}\sqrt{4\frac{m^2\pi^2}{L^2}+\frac{1}{\ell_{x}^2}}w)\right).
\end{eqnarray}
The second term in (\ref{diverge sum}) diverges. This divergence is
a result of the fact that by evaluating the voltage between $x=0$
and $x=L$, we place our voltage probes on top of the contacts. If
the probes are put at a distance $\epsilon$ from the current
contacts, we obtain
\begin{eqnarray}\nonumber\label{non diverge sum}
V_{43}=-\rho_{xx}\sum_{m(odd)}&&I\frac{8L}{m^2\pi^2}\left(\frac{1}{2\ell_x}-\frac{1}{2}
\sqrt{4\frac{m^2\pi^2}{L^2}+\frac{1}{\ell_{x}^2}}\right)\times\\
&&\times\cos(\frac{m\pi\epsilon}{L})=S_2-S_1,
\end{eqnarray}
where we define
\begin{eqnarray}\label{sums defined}\nonumber
&&S_1\equiv\rho_{xx}\sum_{m(odd)}I\frac{8L}{m^2\pi^2}\frac{1}{2\ell_x}\cos(\frac{m\pi\epsilon}{L}),\\\nonumber
&&S_2\equiv\rho_{xx}\sum_{m(odd)}I\frac{8L}{m^2\pi^2}\frac{1}{2}
\sqrt{4\frac{m^2\pi^2}{L^2}+\frac{1}{\ell_{x}^2}}\cos(\frac{m\pi\epsilon}{L}).
\end{eqnarray}
The first sum can be evaluated
\begin{eqnarray}\label{first term}
S_1= \rho_{xx}I(\frac{L}{2\ell_x}-\frac{\abs{\epsilon}}{\ell_x}).
\end{eqnarray}
The square root in $S_2$ may be approximated: for $m\ll m_0$ (with
$m_0\equiv L/\pi\abs{\ell_x}$) we have
$\sqrt{4(m\pi/L)^2+1/\ell_x^2}\cong 1/\abs{\ell_x}$, whereas for
large values of $m$ we have $\sqrt{4(m\pi/L)^2+1/\ell_x^2}\cong
2\pi m/L$. Within that approximation
\begin{eqnarray}\label{eq:voltage y=w}
&&V_{43}=
\rho_{xx}I(L-2\epsilon)\left(\frac{1}{2\abs{\ell_x}}-\frac{1}{2\ell_x}\right)+\\\nonumber
&&\rho_{xx}I\sum_{m(odd)=m_0}^{\infty}\frac{8L}{m^2\pi^2}\left(\frac{m\pi}{L}-\frac{1}{2\abs{\ell_x}}\right)\cos{\frac{m\pi\epsilon}{L}}
\end{eqnarray}

The first term in (\ref{eq:voltage y=w}) is the only term that
depends on the sign of $\ell_x$: when $\ell_x < 0$ it is exactly
equal to $I\Delta\rho_{xy}$, where $\Delta\rho_{xy}$ is now the
change in the Hall resistivity over the length $L-2\epsilon$; when
$\ell_x>0$ Eq. (\ref{eq:voltage y=w}) is hard to evaluate
analytically. For $\epsilon\ll\ell_x$, the second term
logarithmically diverges. This divergence, similar to that of
equation (\ref{diverge sum}), originates from the singular current
density near a point source/sink. If $\epsilon>\ell_x$, the
voltage on the $y=w$ side is much smaller than that on the $y=0$
edge, since most of the current turns to flow on the $y=0$ side.
Its precise $\epsilon$--dependence is hard to extract from
(\ref{eq:voltage y=w}).

To summarize the discussion of current flow along the $\rho_{xy}$
gradient in a finite sample, we note that for $w\gg \ell_x$ the side
along which the current flows is determined by the sign of $\ell_x$
and not by the position of the contacts. Along this side (which is
$y=0$ for $\ell_x>0$ and $y=w$ for $\ell_x<0$),as long as the the
voltage contacts are sufficiently far from the current source and
sink, the voltage drop is determined by the difference in
$\rho_{xy}$ between the voltage probes, and is independent of
$\rho_{xx}$. Along the other edge the voltage drop is very small.

\subsection{Gradient in the $y$-direction}
In the case of a uniform gradient in the $y$ direction, the current
injected at the corners propagates primarily along the edge at
$y=w$. In this case, however, the concentration of the current on
the $y=w$ is a result of its injection at the corners of that side,
and not a result of the density gradient. Indeed, as
\S.\ref{Currentdistribution} shows, if the current is injected
uniform in $y$, it stays so. As a consequence, the current profile
is not exponential, and the voltage drop on the $y=w$ is not
independent of $\rho_{xx}$. This qualitative picture comes out of
Eq. (\ref{full solution}). For  $\partial\rho_{xy}/\partial x=0$ we
have

\begin{eqnarray}
\psi =
\sum_{m=0}^{\infty}&&-I\frac{2m\pi(1-(-1)^{m}e^{-\frac{L}{2\ell_{y}}})}{(\frac{L}{2\ell_{y}})^{2}+m^2\pi^2}\times\\\nonumber
&&\times\
\frac{\sinh(\sqrt{\lambda_{m}}y)}{\sinh(\sqrt{\lambda_{m}}w)}\sin(\frac{m\pi
x}{L})\exp{\frac{x}{2\ell_{y}}}.\label{solution2}
\end{eqnarray}

We first consider the voltage drop along $y=0$. As a first step we
calculate $\frac{\partial \psi}{\partial y}|_{y=0}$:
\begin{eqnarray}
\frac{\partial \psi}{\partial y}|_{y=0} =
\sum_{m=0}^{\infty}-I\frac{2m\pi(1-(-1)^{m}e^{-\frac{L}{2\ell_{y}}})}{(\frac{L}{2\ell_{y}})^{2}+m^2\pi^2}
\times\\\nonumber\times\sin(\frac{m\pi
x}{L})\exp{\frac{x}{2\ell_{y}}}\frac{\sqrt{\lambda_{m}}}{\sinh(\sqrt{\lambda_{m}}w)}.
\end{eqnarray}
Taking the limit of $\ell_y\ll L,w$ we can write
$\sinh(\sqrt{\lambda_{m}}w)=1/2\exp{(\sqrt{\lambda_{m}}w)}$. An overestimate of the sum may be obtained by replacing
$\sqrt{\lambda_{m}}\exp{(-\sqrt{\lambda_{m}}w)}$  by $\sqrt{\lambda_{0}}\exp{(-\sqrt{\lambda_{0}}w)}$.
Since for $0<x<L$
\begin{equation}
\sum_{m=0}^{\infty}\frac{2m\pi(1-(-1)^{m}e^{-\frac{L}{2\ell_y}})}{(\frac{L}{2\ell_y})^{2}+m^2\pi^2}\sin{\frac{m\pi
x}{L}}=\exp{-\frac{x}{2\ell_y}},
\end{equation}
we get that
\begin{eqnarray}\nonumber
\left.\frac{\partial \psi}{\partial y}\right|_{y=0} =
-I\sqrt{\lambda_{0}}\exp{(-\sqrt{\lambda_{0}}w)}=\\-I/2\abs{\ell_y}\exp{(-w/2\abs{\ell_y})},
\end{eqnarray}
leading, when integrated over $x$,  to an exponentially small
voltage drop regardless of the sign of the gradient of $\rho_{xy}$.

At $y=w$, we can expect to have the same voltage drop for both
positive and negative gradients. We note that the expression for
the current in the $x$ direction remains unchanged when replacing
$\ell_y$ with $-\ell_y$, along with replacing $x$ by $L-x$;
therefore, it is enough to consider only the case when $\ell_y$ is
positive. Integrating $j_x$ with respect to $x$ and taking the
limit of $\ell_y\ll L,w$ we find the potential difference between
the points $x_4$ and $x_3$ to be
\begin{widetext}
\begin{eqnarray}V_{43}=
-\rho_{xx}\int_{x_3}^{x_4} j_x dx\cong\sum_{m=0}^{\infty}
\rho_{xx}I\frac{2m\pi
L(1-(-1)^{m}e^{-\frac{L}{2\ell_y}})}{\left(\sqrt{(\frac{L}{2\ell_{y}})^{2}+m^2\pi^2}\right)^3}
\left(\frac{1}{2\ell_y}\sin(\frac{m\pi
x}{L})-\frac{m\pi}{L}\cos(\frac{m\pi
x}{L})\right)\exp{\frac{x}{2\ell_{y}}}\left|_{x_4}^{x_3}\right..
\end{eqnarray}
\end{widetext}
 We calculate the voltage
between $x_4=\epsilon$ and $x_3=L-\epsilon$, in the limit of
$\epsilon \gg \ell_y$. Replacing summation by integration, we
obtain
\begin{eqnarray}\label{eq:phi bessel}
V_{43}\cong
\rho_{xx}I\frac{2}{\pi}\left(\frac{x}{2\ell_y}K_0\left(\frac{x}{2\ell_y}\right)+\frac{x}{2\ell_y}K_1\left(\frac{x}{2\ell_y}\right)-\right.\\
\left.K_0\left(\frac{x}{2\ell_y}\right)\right)
\exp{\frac{x}{2\ell_{y}}}\left|_{\epsilon}^{L-\epsilon}\right..\nonumber
\end{eqnarray}
The leading terms are
\begin{eqnarray}
V_{43} \sim \sqrt{\rho_{xx}\Delta\rho_{xy}^{\emph{{\tiny
L}}-\epsilon}}I-\sqrt{\rho_{xx}\Delta\rho_{xy}^{\epsilon}}I
\end{eqnarray}
where
\begin{eqnarray}\nonumber
&&\Delta\rho_{xy}^{\emph{{\tiny
L}}-\epsilon} =\frac{\partial\rho_{xy}}{\partial y} (L-\epsilon),
\\
&&\Delta\rho_{xy}^{\epsilon}=\frac{\partial\rho_{xy}}{\partial
y}{\epsilon}\nonumber.
\end{eqnarray}
This time, the voltage explicitly depends on $\rho_{xx}$, as the width of the strip within which the current flows does not
scale as $\rho_{xx}^{-1}$.

\section{Uniform Dissipative Region between two Quantum Hall regions}

\label{sec:dissipative}

We now consider in detail a situation in which there are quantized
Hall states at the two ends of the sample as in Fig.~\ref{contour},
with Hall resistivities ${\rho}_{xy}^{A}$ and ${\rho}_{xy}^{B}$ and
vanishing longitudinal resistivity. In between there must be a
dissipative region, where the local dissipative resistivity
$\rho_{xx}$ is not necessarily small compared to $\rho_{xy}$.  In
general, the resistivity in this region will be $x$--dependent.
However, we simplify and consider the central region as having
longitudinal resistivity ${\rho}_{xx}^{C}$ and Hall resistivity
$\rho_{xy}^{C}$ that do not vary in position (see Fig.~\ref{system}).

Since the resistivity components vary along the $x$ direction only, Eq. (\ref{fullcurl})
becomes
\begin{equation}
\rho_{xx}(x)\nabla^2\psi+\rho_{xy}'(x)\frac{\partial\psi}{\partial
y} +\rho_{xx}'(x)\frac{\partial\psi}{\partial x}=0
\label{fullcurl2}
\end{equation}
where $\rho_{ij}'\equiv\frac{\partial\rho_{ij}}{\partial x}$. Thus,
in the central region where $\rho_{ij}'=0$, the stream function
$\psi$ simply obeys the Laplace equation.  The interfaces to the
quantized Hall regions $A$ and $B$ impose boundary conditions on the
current density (and hence on $\psi$). Conservation of the electric
field parallel to the interface and of the current density
perpendicular to the interface, leads to
\begin{eqnarray} \label{boundaryX}
\frac{{j}_x}{{j}_y}=\left\{
\begin{array}{ll}
-\frac{{\rho}_{xx}^{C}}{\rho_{xy}^{A}-\rho_{xy}^C} & \quad x=0, \\
\\
-\frac{\rho_{xx}^{C}}{{\rho}_{xy}^{B}-\rho_{xy}^C} & \quad x=L, \\
\end{array}
\right.
\end{eqnarray}
where $j_{x}=-\partial \psi/\partial y$, $j_{y}= \partial
\psi/\partial x$ are the components of the current density inside
the central dissipative region. At the $y=0$ and $y=w$ edges of the
sample the current must be parallel to the edge, {\it i.e.}
\begin{equation}
j_y(y=0)=j_y(y=w)=0. \label{boundaryY}
\end{equation}

We simplify matters by using a conformal mapping\cite{RG}.  We
introduce a complex potential
\begin{equation}
V=\chi+\imath\psi
 \label{complex potential}
\end{equation}
where $\psi$ is the stream function. The function $V$ is analytic
in $z=x+\imath y$ and hence $\psi$ obeys the Laplace equation.
Using the Cauchy-Reimann equations we may find the full derivative
of $V$ with respect to $z$
\begin{equation}
\frac{dV}{dz}=-j_x+\imath j_y .\label{complexfield}
\end{equation}
So the full derivative of the complex potential is a linear
combination of the components of the current density. If we define
\begin{equation}
\frac{dV}{dz}=e^{f(z)} \label{exponent}
\end{equation}
where $f(z)$ is an analytic function, we get that
\begin{eqnarray}
&&j_x=-e^{{\rm Re}(f(z))}\cos {\rm Im}(f(z))\label{jx},\\
&&j_y=e^{{\rm Re}(f(z))}\sin {\rm
Im}(f(z))\label{Ey}.
\end{eqnarray}
The boundary conditions set for the stream function $\psi$ can now
be translated into conditions on the function $f(z)$. The
requirement on setting $j_y$ to zero at $y=(0,w)$ is translated
into setting the imaginary part of $f(z)$ to zero there. The
boundary conditions at $x=(0,L)$ can be translated into
\begin{eqnarray} \label{angle}
\tan{\rm Im} f(z)=\left\{
\begin{array}{ll}
\frac{{\rho}_{xy}^{A}-\rho_{xy}^C}{\rho_{xx}^C} & \quad x = 0,
\\\\
\frac{{\rho}_{xy}^B-\rho_{xy}^C}{\rho_{xx}^C} & \quad x = L. \\
\end{array}
\right.
\end{eqnarray}

Once we have solved for the imaginary part of $f(z)$, $f(z)$ is
completely determined, and therefore the electric field and current
distribution are known up to a normalization factor, which
determines the current direction and magnitude. We find
\begin{eqnarray}\label{FofZ}
f(z)&=&\sum_{n(odd)=1}^{\infty}\frac{4}{n\pi}\left(\sinh\frac{n\pi
L}{2w}\right)^{-1}\times\\&&\times\left(\beta_B\cosh\frac{n\pi
z}{w}-\beta_A\cosh\frac{n\pi (z-L)}{w}\right)\nonumber
\end{eqnarray}
where
$\beta_A\equiv\arctan[({\rho}_{xy}^A-\rho_{xy}^C)/\rho_{xx}^C]$,
$\beta_B\equiv\arctan[({\rho}_{xy}^B-\rho_{xy}^C)/\rho_{xx}^C]$.
Since the intermediate region is a transition region between the
two quantized Hall states on its ends, ${\rho}_{xy}^C$ is
intermediate between ${\rho}_{xy}^A$ and ${\rho}_{xy}^B$. Thus,
one of the numbers $\beta_A$ and $\beta_B$ is negative and the
other is positive. We set $\beta_A<0$ and $\beta_B>0$. Evaluating
$f(z)$ on the boundary at $x=L$ we get:
\begin{eqnarray}
&&{\rm Im}f(x=
L,y)=\sum_{n(odd)=1}^{\infty}\frac{4\beta_B}{n\pi}\sin\frac{n\pi
y}{w} =\beta_B\nonumber,\\\nonumber &&{\rm Re}f(x=
L,y)=\sum_{n(odd)=1}^{\infty}\frac{4}{n\pi}\cos{\frac{n\pi
y}{w}}\times\\&&\quad\quad\quad\quad\times\left(\beta_B\coth{\frac{n\pi
L}{w}}-\frac{\beta_A}{\sinh{\frac{n\pi L}{w}}}\right).
\end{eqnarray}
At the corners $y=(0,w)$ the real part of $f(z)$ diverges, going to
$+ \infty$ at $y=0$, and $-\infty$ at $y=w$. This divergence implies
the current density diverges at the corner $(L,0)$ of the
dissipative region, and vanishes at $(L,w)$. The same behaviour is
observed at $x = 0$, implying that current enters and leaves the
dissipative region from the corners. Interchanging the signs of
$\beta_A$ and $\beta_B$ moves the points of divergence to the line
$y=w$.

In the limit where $w\gg L$, for every $x$ in the range $0<x<L$ we
get
\begin{equation}
{\rm
Re}f(x,y)\simeq\\\sum_{n(odd)=1}^{\infty}\frac{4(\beta_B-\beta_A) w}{\pi^2 L}\frac{\cos\frac{n\pi
y}{w}}{n^2}.
\end{equation}
This sum can be carried out explicitly to obtain
\begin{equation}
{\rm Re}f(x,y)\simeq\frac{(\beta_B-\beta_A)}{L}\left(\frac{w}{2}-\abs{y}\right).\label{realFapprox}
\end{equation}
implying the current density to be large near the edge at $y=0$,
and decay exponentially as a function of $y$. The decay length is
$\ell=L/(\beta_B-\beta_A)$. In the limit of small dissipation
where $\rho_{xx}^C\ll |{\rho}_{xy}^{A,B}-\rho_{xy}^C|$ we have
$\abs{\beta_{A,B}}\rightarrow\pi/2$ and the decay length
approaches the limit $\ell=L/\pi$.  In the opposite limit of large
dissipation where ${\rho}_{xx}^{C}\gg |{\rho}_{xy}^{A,B} -
\rho_{xy}^C|$, the decay length grows like
$\rho_{xx}^C/|{\rho}_{xy}^A-{\rho}_{xy}^{B}|$. As long as this
length is smaller than $w$, the current is still concentrated
along one side of the sample, and the longitudinal voltage on that
side is determined by $\rho_{xy}^{A}-\rho_{xy}^{B}$, rather than
by $\rho_{xx}$.  Again, interchanging the sign of $\beta_{A}$ and
$\beta_{B}$ switches the edge where the current propagates from
$y=0$ to $y=w$.
\begin{figure}[t]
\begin{center}
    \includegraphics[clip,width=70mm]{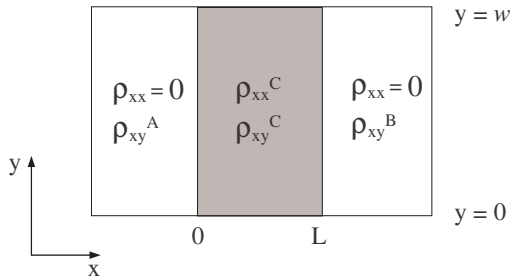}
    \caption{
The three different regions of the system described in
\S~\ref{sec:dissipative}. } \label{system}
\end{center}
\end{figure}

\section{Discussion}

\label{sec:discuss}

In this work we discussed the effect of density
inhomogeneities on the measurement of the longitudinal resistance
in quantum Hall systems.

We analyzed the case in which the two ends of the sample are in two
different quantized Hall states, separated by a dissipative transition
region. Generically, in this situation we find the current flow to be
concentrated along one of the edges. We showed that when that happens,
the voltage along that edge is controlled by the difference in Hall
resistivities between the voltage probes, while the voltage on the
other edge is very small. These observations, which are largely
independent of the specific spatial dependence of both $\rho_{xx}$ and
$\rho_{xy}$ in the dissipative region, correspond to what is observed
in the experiment. We solved in detail two simplified models, which
indeed showed a concentration of the current along one of the edges.

In one of the models we solved, the longitudinal resistivity is
assumed independent of position, and the Hall resistivity has a
uniform gradient. In such a case, the current behavior is controlled
by the two length scales $\ell_x$ and $\ell_y$. For samples large
compared to these length scales, the current flows in a very
asymmetric way through the sample. This has been illustrated by
explicit calculations in rectangular geometries, both for the ideal
case of a long system, and for a rectangular system where the location
of the contacts has to be taken into account. We have shown that in
both cases, when the gradient is in the direction along which the
current flows, the longitudinal resistance will be
$\Delta\rho_{xy}$ on one side of the sample, and much smaller on the
other. For the long system, we have shown that this result still
holds, even if in addition there is a transverse component to the
gradient.

For a long Hall bar geometry with a linear gradient in the local
Hall resistivity, we showed how the microscopic resistivity
$\rho_{xx}$ can be extracted from the measurements of the
longitudinal voltage drop on the two sides of the sample, via
Eq.~(\ref{eq:rhoxx}).  We believe that this model can be of
application to the cases of transport measurements on compressible
states in the quantum Hall regime, for example the $\nu=1/2$
composite-fermion liquid state in high quality samples.
The conditions for application of this model to experiment relies on
the validity of the assumptions used to derive
Eq.~(\ref{eq:simple}).
Firstly, these assumptions require that the variation in $\rho_{xx}$
over the size of the sample is small compared to its average value;
typically this means that the fractional change in carrier density
should be small.  (We note, however, that for a system with small
dissipation, $\rho_{xx} \ll \left|\rho_{xy}\right|$ -- that is with a
local Hall angle close to $90^\circ$ -- even a small fractional change
in carrier density can allow the change in the Hall resistivity
$\rho_{xy}$ to be large compared to $\rho_{xx}$.)
Secondly, we have assumed that we can retain only the linear
gradient in the local Hall resistivity $\rho_{xy}(\mathbf{r})$.
The validity of this assumption depends on the functional form of
$\rho_{xy}(\mathbf{r})$. In the case where variations in
$\rho_{xy}(\mathbf{r})$ are all on the same characteristic length
scale, which we shall refer to as the macroscopic correlation
length $\xi_M$, this assumption is valid provided $\xi_M \gg L,
\mbox{min}(w,\ell_x)$. In practice, $\xi_M$ can be very large, set
by inhomogeneities in the growth of the samples related to
characteristics of MBE machines. In realistic high mobility 2DEGs,
one anticipates that there are additional inhomogeneities on a
microscopic lengthscale, $\xi_m$, of order the spacer-layer
thickness. The model remains valid provided this additional
lengthscale is sufficiently small, $\xi_m \ll L,
\mbox{min}(w,\ell_x)$, that these microscopic inhomogeneities
simply affect the effective local resistivity tensor which varies
smoothly on the longer lengthscale $\xi_M$. \vspace{-0.5cm}
\acknowledgements{This research was partly supported by the UK
EPSRC Grant No. GR/R99027/01 (N.R.C.). AS and RI thank the
US-Israel Bi-National Science Foundation, the Israel Science
Foundation and the Minerva foundation for financial support. }

\appendix*
\section{Calculation of the dissipated power}

\label{sec:appendix}

Using the same approach as in Ref.~\onlinecite{rch}, we derive
expressions for the power dissipation in a sample with quantized
Hall states $A$ and $B$ at the two ends, as in Fig.~\ref{contour}.
The central region has arbitrary local resistivity tensor.

The power dissipation is given by
\begin{eqnarray}
Q=\int\int
\mathbf{E}(\mathbf{r})\cdot\mathbf{j}(\mathbf{r})\;d^{2}\mathbf{r}\nonumber.
\end{eqnarray}
Since in regions $A$ and $B$ the resistivity tensor has only
off-diagonal elements which are non zero, in these two regions
equations (\ref{stream}) (\ref{static}) and (\ref{rho}) can be
combined into a single equation,
\begin{eqnarray}
\rho\mathbf{\nabla}\psi=-\mathbf{\nabla}\phi,
\end{eqnarray}
where $\rho$ is a scalar, equal to $\rho_{xy}^{A},\rho_{xy}^{B}$
in regions $A$ and $B$. This equation can now be
integrated to obtain the relation between $\phi$ and $\psi$:
\begin{eqnarray}\label{relation_scalars}
\phi=-\rho\psi+c,
\end{eqnarray}
where $c$ is a constant, in principle different for regions $A$
and $B$. Writing $Q$ in terms of these scalar functions and using
Stokes' theorem, the dissipation integral becomes
\begin{eqnarray}
Q=-\int(\psi\mathbf{\nabla}\phi)\cdot d\mathbf{l}.
\end{eqnarray}
We now integrate over a rectangle that encloses the central region,
as demonstrated in Fig.~\ref{contour}. We note that at $y=(0,w)$,
the stream function $\psi$ must be a constant, since we do not allow
current to flow in the $y$ direction. We denote $\psi(y=0)\equiv
\psi_{b}$ and $\psi(y=w)\equiv\psi_{t}$, with
\begin{eqnarray}\label{Total Current}
I=\psi_{b}-\psi_{t}
\end{eqnarray}
being the total current  flowing from left to right. Performing
the integration we get
\begin{eqnarray}\label{eq:q2}\nonumber
Q =
-\psi_b(\phi_2-\phi_1)-\psi_t(\phi_4-\phi_3)+\\\frac{1}{2}(\psi_b^2-\psi_t^2)(\rho_{xy}^A-\rho_{xy}^B).
\end{eqnarray}
Using (\ref{relation_scalars},\ref{Total Current}) we can write
\begin{eqnarray}\label{dissipation2}
Q&=&\frac{1}{2}I^{2}(\rho_{xy}^{A}-\rho_{xy}^{B})+I(\phi_{1}-\phi_{2})
\\\label{dissipation2p}
&=&\frac{1}{2}I^{2}(\rho_{xy}^{B}-\rho_{xy}^{A})+I(\phi_{4}-\phi_{3}).
\end{eqnarray}
Note that in \S\ref{sec:general} we argue that the potential drop
along one side of the sample can become exponentially small.
Clearly, from (\ref{dissipation2}) if $V_{12}= \phi_1-\phi_2$ is
to be vanishingly small, then we must have
$\rho^A_{xy}-\rho_{xy}^B > 0$ such that the power loss is
positive; similarly from (\ref{dissipation2p}) $V_{43}=
\phi_4-\phi_3$ can be vanishingly small only if
$\rho^A_{xy}-\rho_{xy}^B < 0$.

The general result can be stated most clearly by noting that, using
(\ref{relation_scalars},\ref{Total Current}), we can express the
current in terms of voltages as
\begin{eqnarray}
I =  \frac{V_{41}-V_{32}}{\rho_{xy}^A-\rho_{xy}^B} =
\frac{V_{43}-V_{12}}{\rho_{xy}^A-\rho_{xy}^B}.
\end{eqnarray}
Substituting this in (\ref{dissipation2}) one then obtains
\begin{eqnarray}
Q = \frac{1}{2}(\rho_{xy}^A-\rho_{xy}^B)^{-1}(V_{43}^2-V_{12}^2),
\end{eqnarray}
which indicates that $|V_{43}| > |V_{12}|$ for $\rho_{xy}^A>
\rho_{xy}^B$, and vice versa, as discussed in \S\ref{sec:general}.
%

\end{document}